\documentclass[fp,onecolumn]{jpsj3}
\usepackage{txfonts}
\usepackage{graphicx}

\title{Precision Muonium Spectroscopy}

\author{Klaus P. Jungmann}
\inst{Van Swinderen Institute, University of Groningen, Zernikelaan 25, 9747 AA Groningen, The Netherlands}

\abst{The muonium atom is the purely leptonic bound state of a positive muon and an electron. It has a lifetime of 2.2 $\mu$s. The absence of any known internal structure provides for  precision experiments to test fundamental physics theories and to determine accurate values of fundamental constants. In particular groun dstate hyperfine structure transitions can be measured by microwave spectroscopy to deliver the muon magnetic moment. The frequency of the  1s-2s transition in the hydrogen-like atom can be determined with laser spectroscopy to obtain the muon mass. With such measurements fundamental physical interactions, in particular Quantum Electrodynamics, can also be tested at highest precision. The results are important input parameters for experiments on the muon magnetic anomaly. The simplicity of  the atom enables further precise experiments, such as a search for muonium-antimuonium conversion for testing charged lepton number conservation and searches for possible antigravity of muons and dark matter.} 
\begin{document}

\maketitle

\section{Introduction}
Muonium (M = $(\mu^+e^-)$) has been discovered by Hughes and collaborators in 1960 \cite{Hughes_1960,Telegdi_1991,Hughes_2000}, some 26 years after the first muon had appeared in a Wilson cloud chamber \cite{Kuntze_1934}. M consists of two leptons from two different particle generations. The particles are bound by the electromagnetic force. The positive muon ($\mu^+$) is an antiparticle from the second generation. The electron ($e^-$) is the lightest charged lepton and from the first generation. Leptons are not directly subject to the strong force and they are {\it point-like}, i.e.,  to the best of our knowledge they have no inner structure and all their known structure arises from well calculable virtual fields in their vicinity \cite{Kinoshita_1990}.\\ 
The lifetime $\tau_{M}$ of the exotic atom is essentially given by the free muon lifetime $\tau_{\mu}$\cite{Webber_2011}.
With the presently achievable precision they differ insignificantly by relative $\approx 6\cdot 10^{-10}$,
which is due to time dilation in the bound state \cite{Czarnecki_2000}.   The  lifetime,
\begin{equation}
\tau_{M}= \tau_{\mu} = 2 .196\,980\,3(2\,2)~\mu\rm{s} ~~,
\end{equation}
is sufficiently long to enable a number of precision experiments in various fields of physics, including atomic, particle and general physics as well as applications in, e.g., condensed matter and life sciences \cite{Lee_1999}. \\
Precision  experiments on M, such as on the  hyperfine splitting $\Delta \nu_{HFS}$, its Zeeman effect, the 1s-2s frequency difference $\Delta \nu_{1s2s}$ and a search for spontaneous muonium to antimuonium (M$\overline{\rm M}$) conversion, could not be performed in this millennium, yet. All the measurements, which have been conducted to date, are essentially statistics limited \cite{Jungmann_2006}. They had in their precision stage a necessity for long, extended beam time periods. For a number of such precision experiments on M the intense $\mu^+$ beams at the Japan Proton Accelerator Research Complex (J-PARC), Tokai, Japan, have opened new and important possibilities for which the community has been waiting \cite{Aysto_2001,Jungmann_2001,Bandyopadhyay_2009}. Refined precision experiments on M  and muonic atoms are needed in particular, because of the puzzling results reported from muonic hydrogen 
($\mu^-$H) spectroscopy \cite{Pohl_2010,Pohl_2013,Antognini_2013}, where the extracted value for the proton charge radius differs from values obtained by electron scattering. An explanation is still pending.\\
The success of ongoing research concerning the muon magnetic anomaly $a_{\mu}$ in g-2 experiments at Fermi National Accelerator center (FNAL), Batavia, U.S.A.,\cite{Grange_2015} and at J-PARC \cite{Mibe_2011,Saito_2012} will depend on reliable values of fundamental constants \cite{Mohr_2012}, in particular on the muon magnetic moment $\mu_{\mu}$ and the muon mass $m_{\mu}$. These parameters are related through 
\begin {equation}
a_{\mu}=\frac{\omega_a m_{\mu}c}{e_{\mu} B} = \frac{\frac{\omega_a}{\omega_p}}{\frac{\mu_{\mu}}{\mu_p}-\frac{\omega_a}{\omega_p}},
\end{equation}
where $e_{\mu}$ is the muon electric charge, $c$ the speed of light, $B$ the magnetic field in which the experiment takes place, $\omega_a$ the anomaly frequency measured in a muon g-2 experiment, $\mu_p$ the proton magnetic moment and $\omega_p$ the proton NMR frequency, if $B$ is  measured using proton NMR, typically in water. In a system of measurements the influence of  imperfections in the magnetic field measurement devices can be arranged to largely cancel, if both $a_{\mu}$ and $\mu_{\mu}$ are determined with the magnetic field measured by the very same devices and procedures. Such an arrangement, which is limited in principle by statistical uncertainties and external factors affecting reproducibility only, has been in  in place for the latest muon g-2 experiment at Brookhaven National Laboratory (BNL), Upton, U.S.A., and the measurements of the Zeeman effect of the ground state hyperfine structure (HFS) in M at the  Los Alamos Meson Physics Facility (LAMPF), Los Alamos, U.S.A. The experiments shared the very same field measurement system based on pulsed proton NMR in water, which provides for field measurements at absolute precision $3.4\cdot 10^{-8}$  \cite{Prigl_1996,Fei_1997}. \\
For a new round of precision experiments improvement of the $B$ field measurement can be achieved with optically pumped, hyperpolarized $^3$He \cite{Neumayer_1999,Nikiel_2014}. In this case the sensitivity to the NMR sample container shape would be decreased by some three orders of magnitude, reflecting the density ratio of the NMR samples.\\
Besides spectroscopy on M  there is a particularly rich field of fundamental precision measurements on simple muonic atoms. They
have potential to advance QED, electroweak physics and nuclear physics \cite{Jungmann_1992,Boshier_1996}. Precision laser and microwave spectroscopy experiments on M, ($\mu^{-4}$He), $(\mu^{-3}$He), and  $(\mu^-$H) are underway and already produced puzzling results, e.g., concerning the extracted proton (p) radius\cite{Pohl_2010}. Furthermore, in the past decade new and novel experimental research has been suggested beyond optical and microwave spectroscopy conducted to date.\\ 

\section{Theory}
M can be considered a light hydrogen (H) isotope.
With the masses of $\mu^+$ and $e^-$ differing significantly, the theoretical treatment of  M  is analogous to atomic H, except nuclear structure is absent in M.
E.g., in H this structure limits the comparison of experiment and theory for the HFS in the n=1 state at relative precision ppm. While experiments achieved relative uncertainty $10^{-13}$, theory is hindered by the knowledge of nuclear structure and  polarizability \cite{Sapirstein_1990}. In view of some 7 orders of magnitude difference in accuracy, little progress has been made in the past 3 decades. The M atom consists exclusively of leptons without inner structure, a comparison of theory and experiment concerning $\Delta \nu_{HFS}$ is possible at one order of magnitude higher precision than for H. The limits on a comparison arise here from the knowledge of m$_{\mu^+}$ and $\mu_{\mu}$, respectively, which in reverse both can be improved by more precise measurements.\\
\begin{figure}
\vspace*{7mm}
\hspace*{4mm}
 \includegraphics[width=7.5cm]{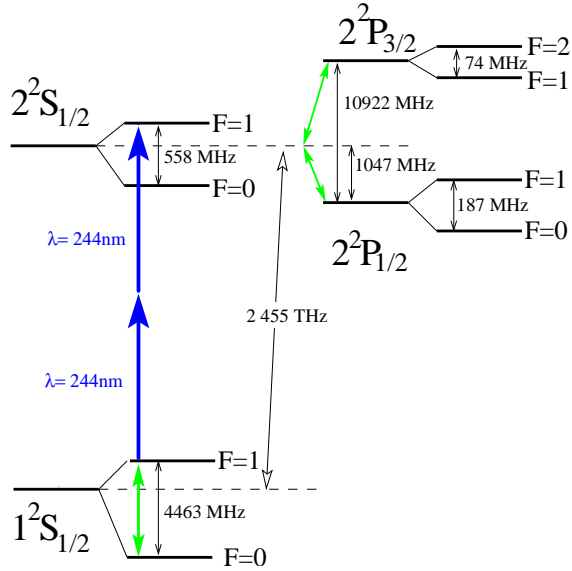}
\vspace*{5mm}
 \caption{Muonium atom n=1 and n=2 energy levels \cite{Jungmann_2001}. All indicated transitions have been observed. The ground state hyperfine interval and the 1$^2$S$_{1/2}$,F=1-2$^2$S$_{1/2}$,F=1 transition have been measured and calculated very precisely. The transitions within the excited states could be demonstrated.}
 \label{81403Fig1}
\vspace*{-5.6mm}
\end{figure}
The level energies of the isotopes hydrogen H, deuterium D, tritium T, as well as  M (Fig. \ref{81403Fig1}), are given by
 \cite{Bethe_1957,Sapirstein_1990}
\begin{eqnarray}
E_{tot}(n,j,l,F)& = &E_D(n,j) + E_{RM}(n,j,l) + E_{QED}(n,j,l) \nonumber \\
		& + &E_{HFS}(n,j,l,F,I) + E_{weak} + E_{exotic} \hspace*{5mm},
\end{eqnarray}
with $n$ the principal quantum number, $j$ the electron angular momentum, the
$l$ the orbital angular momentum, $F$ the total angular momentum, and $I$ the
{\it nuclear} Spin.
The dominating term is the Dirac energy
\begin{equation}
E_D(n,j)= m_e c^2
    \left( f(n,j) - 1
    \right)~~,
\end{equation}
where
\begin{equation}
f(n,j) =
    \sqrt{\frac{1}{1+
    \left( \frac{Z\alpha}{n-\varepsilon}
    \right) ^2}}~,
%\end{equation}
%with 
%\begin{equation}
\varepsilon =
    j+\frac{1}{2} - \sqrt{
    \left(  j+\frac{1}{2}
    \right) ^2-(Z\alpha)^2}~.
\end{equation}\\
Dirac theory describes the system to the fine structure level.
A reduced mass term $E_{RM}$ accounts for finite nuclear mass.
This contribution has two parts,
\begin{equation}
\label{RM}
E_{RM}(n,j,l) = E_{NRRM} + E_{RRM}(n,j,l)~~.
\end{equation}
$E_{NRRM}$ describes  the classical nonrelativistic reduced mass %term
\begin{equation}
E_{NRRM}=
    \left( \frac{m_r}{m_e}-1
    \right) \cdot E_D \\
\end{equation}
and $ E_{RRM}(n,j,l)$ is the relativistic reduced mass effect which arises from full relativistic treatment of the two-body problem,  
\begin{eqnarray}
E_{RRM}(n,j,l) & = & -\frac{m_r^2c^2}{2(m_e+m_N)}  \left(  f(n,j)-1 \right) ^2  \nonumber \\
               & & \hspace{-1.5cm}+\frac{(Z\alpha)^4m_r^3c^2}{2n^3m_N^2} \left(  \frac{1}{j+\frac{1}{2}} -\frac{1}{l+\frac{1}{2}} \right) (1-\delta_{l0})~~.\hspace*{5mm}
\end{eqnarray}
Here $l$ is the orbital angular momentum, 
$m_r= m_em_N/(m_e+m_N)$ the reduced mass,
$m_N$ the {\it nuclear} mass, and 
$\alpha$ the fine structure constant.
Higher order terms are neglected.\\
Additional corrections are due to QED effects,
 the hyperfine interaction ($E_{HFS}(n,j,l,F)$),
 the weak interaction ($E_{weak}$),
 and possibly from exotic processes ($E_{exotic}$), which are 
 not yet provided in the Standard Model.
The QED corrections to the gross structure in M have been subject to constant refinement % over the years 
and they are very similar to natural atomic H and its isotopes. The QED contributions consist of one- and two-loop corrections, relativistic and radiative recoil corrections. 
%There contributions from weak interactions (see eq. \ref{weak}) and possible exotic interactions. 
For M there are no nuclear finite size effects.\\
The interaction of the magnetic moment $\mu_N$ of the nucleus with spin I
and the electron magnetic moment $\mu_e$ causes HFS.
To good approximation the corresponding contribution to the level energies is \cite{Fermi_1930}
\begin{eqnarray}
E_{HFS}(n,j,l,F,I)& = & (Z\alpha)^2 Z g_I \left[\frac{m_e}{m_N} (1+\varepsilon_{{\small QED}}) \left( \frac{m_r}{m_e}\right)^3 \right. \nonumber \\ 
&& \left. \hspace*{-2cm}\times \hspace*{1mm}\frac{hcR_{\infty}} {n^3} \frac{F(F+1)-I(I+1)-j(j+1)}{j(j+1)(2l+1)}\right]~~.
\end{eqnarray}
where $R_{\infty}$ is the Rydberg constant and
$\varepsilon_{QED}$ takes into account relativistic effects,
radiative and recoil corrections as well as the finite nuclear size and nuclear
polarizability,
\begin{equation}
\varepsilon_{QED}   = \varepsilon_{rad}
		    + \varepsilon_{rec}
		    + \varepsilon_{rad-rec}
		    + \varepsilon_{nucl-size}
		    + \varepsilon_{nucl-pol}~~.
\end{equation}
For the M atom, where nuclear structure effects are absent,
the
ground state splitting between the F=0 and F=1 levels
can be expressed as 
\begin{eqnarray}
 \Delta \nu_{HFS}&  =   &  \left[  (Z \alpha)^2
		     R_{\infty} \frac{\mu_{\mu}}{\mu_B}\left( 1 + \frac{m_e}{m_{\mu}} \right) ^{-3}
		    \right.\nonumber \\
		  && \left. \hspace*{-2.5cm} \times \frac{16}{3} \left( 1+ \varepsilon_{rad}+ \varepsilon_{rec}+
		    \varepsilon_{rad-rec} \right) +
		    \Delta \nu_{weak} +\Delta \nu_{exotic} \right]~.
\end{eqnarray}
The QED corrections to $\Delta \nu_{HFS}$ include radiative, recoil and combined radiative recoil corrections. The calculations have been refined recently \cite{Eides_2014}. Weak interactions \cite{Eides_1996} and possible exotic effects have been included.
%%%
\begin{figure}
\vspace*{4.9 cm}
 \includegraphics[width=8.5cm]{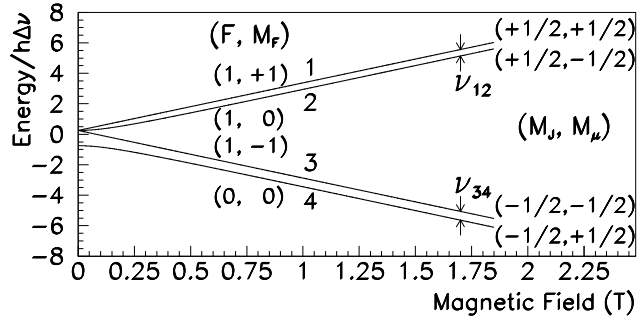}
\vspace*{-4.2cm}
 \caption{The Zeeman sublevels of the n=1 state in M \cite{Jungmann_2004}. The indicated transitions $\nu_{12}$ and $\nu_{34}$ are induced by microwaves and correspond to a flip of the spin of the  $\mu^+$. The transition can be observed through a change in the $\mu^+$ decay asymmetry. The sum of both transition frequencies equals the zero field splitting $\Delta \nu_{HFS}$ and their difference yields the muon magnetic moment $\mu_{\mu}$, if the magnetic field is known well enough.}
 \label{81403Fig2}
\end{figure}
The weak interactions arise from Z boson exchange and yield
a contribution to $E_{tot}(n,j,l,F)$. Due to the short
range of  weak interactions associated with the Z boson mass
($\approx$91~GeV$/c^2$) they can be modeled as a
point current-current interaction with an effective vector-axial vector (V-A) type Hamiltonian.
In the M atom with the coupling between $e^-$ and $\mu^+$ the energy of s-states is shifted by
\begin{equation}
\label{weak}
\Delta E_{weak}^{M}(nS)= (\frac{1}{n^3})
    \frac{\sqrt{2}G_F m_e^2 \alpha^2 m_e c^2}{\pi^2} C_{e\mu}^{VV}
	=\frac{7.2}{n^3}(C_{e\mu}^{VV})~{\rm kHz}~,
\end{equation}
where $G_F$ is the Fermi weak interaction coupling constant and $C_{e\mu}^{VV} = -1/2(1-4\cdot \sin^2 \Theta_W)^2$ is the
vector-vector coupling constant between $e^-$ and $\mu^-$, $\Theta_W$ is the weak mixing angle.
This results for M in a contribution to the 1S-2S  separation of
24~Hz, which is far below the resolution for the foreseeable future.
For H the corresponding shift is 190~Hz; it is significantly larger, because of the difference in  
the axialvector - axialvector coupling contribution to $\Delta \nu_{HFS}$. 
%and the axial current is proportional to the {\it nuclear} spin. 
The hyperfine splitting of H and of M are affected differently by the weak interactions due to the different  muon - electron and nucleus - electron couplings.
For M we have  $C_{e\mu}^{AA}=1/2$ and the parity conserving weak effect on $\Delta \nu_{HFS}$ is \cite{Eides_1996}
\begin{equation}
 \delta \nu _{HFS,weak}^M
    = \frac{2\sqrt{2}G_F m_e^2 \alpha^2 m_e c^2}{\pi^2}
			\cdot C_{e\mu}^{AA}~
%    = -140 \cdot C_{e\mu}^{AA}~{\rm Hz}~
=~-65~{\rm Hz}~.
\end{equation}
As the  $\mu^+$ is an antiparticle there are opposite signs for the effect in M and in H, because the latter consists of particles only.
The size of $ \delta \nu _{HFS,weak}^M$  is about the size of the uncertainty in the most recent measurement of $\Delta \nu _{HFS}$ at LAMPF \cite{Liu_1999}.\\
Potential M$\overline{\rm M}$ oscillations would influence the energy levels in M. 
The ground state of the  coupled M$\overline{\rm M}$ system has eight
energy eigenstates which are different in energy from the four eigenstates
of the uncoupled M and $\overline{\rm M}$ atoms.
In the absence of external $B$ fields the M$\overline{\rm M}$ mixing causes for s state HFS levels a splitting
\begin{equation}
\delta \nu_{M\overline{M}}(nS) =
\langle M | H_{M\overline{M}} | \overline{M} \rangle =
 \frac{519}{n^3} \cdot (G_{M\overline{M}}/G_F) ~{\rm Hz}~~,
\end{equation}
where $G_{M\overline{M}}$ is the coupling constant in an effective four fermion interaction and $G_F$ the Fermi coupling constant of the weak interactions \cite{Schafer_1988}.
With the present limit on $G_{M\overline{M}}/G_F$ \cite{Willmann_1999} we have
$\delta \nu_{M\overline{M}}(1S)\le~1.5$~Hz in the n=1 state, 
which is below the present precision goal for theory and near future possible experiments,
in particular since  M$\overline{\rm M}$ conversion is suppressed in 
external magnetic fields \cite{Horikawa_1996}.\\
Further possibilities to search for new effects, e.g. dark forces, have been investigated. The n=1 state HFS  measurements appear to provide
here already nontrivial constraints on such forces\cite{Karshenboim_2014}.
\\
The Hamiltonian describing the n=1 levels of M
in an external magnetic field ${\bf B}$ is \cite{Hughes_1977}
\begin{equation}
\label{HAMN1}
H_Z = A_{HFS} {\bf I_{\mu} \cdot J} + \mu_B g_J {\bf J \cdot B}
- \frac{m_e}{m_{\mu}} \mu_B g'_{\mu}{\bf I_{\mu} \cdot B}~~,
\end{equation}
where $A_{HFS}$ is the n=1 state HFS energy interval
$h\Delta \nu_{HFS}$,
${\bf I_{\mu}}$ is the muon spin operator, ${\bf J}$ is the electron total
angular momentum, $\mu_B$ is the Bohr magnetron,
 and $g_J$ and $g'_{\mu}$
are the gyromagnetic ratios of the $e^-$ and
the $\mu^+$ in M atom. The latter differ
from the free particle  values  $g_e$ and $g_{\mu}$
due to relativistic binding corrections \cite{Grotch_1971},
\begin{equation}
g_{\mu}`  = g_{\mu}
    \left[1-\frac{\alpha^2}{3}+\frac{\alpha^2}{2}\frac{m_e}{m_{\mu}}
    \right],
g_{J}   =   g_{e}
    \left[1-\frac{\alpha^2}{3}+\frac{\alpha^2}{2}\frac{m_e}{m_{\mu}}
	+\frac{\alpha^3}{4\pi}
    \right].		
\end{equation} 
The behaviour of the magnetic sublevels of the F=1 and F=0  hyperfine
states in an external magnetic field ${\bf B}$ (see Fig.\ref{81403Fig2})\cite{Jungmann_2004}
can be 
expressed as \cite{Hughes_1977}
\begin{eqnarray}
\label{ZEEMAN}
\hspace*{-5mm} E(n{\rm=}1,F,M_F)& = &    - g_{\mu'}(m_e/m_{\mu})\mu_BM_F B
                      -\frac{h A_{HFS}}{4}                         \nonumber \\
&&  \hspace*{-1cm}  -(-1)^F \frac{h A_{HFS}}{2} \sqrt{1+2M_Fx+x^2}~~,
\end{eqnarray}
where
\begin{equation}
x = (g_J+g'_{\mu}(m_e/m_{\mu}))\mu_B B / (h ~A_{HFS})
\label{XPAR}
\end{equation}
is the magnetic field parameter, and
$ x=1 $ corresponds to magnetic field $B\approx 0.1585~{\rm T}$. 
The Breit-Rabi equation (eq. \ref{ZEEMAN}) does not incorporate the effect of higher quantum
states. Such corrections of relative order $(\Delta
\nu_{HFS}/R_{\infty})^2\approx 10^{-12}$ are negligible. Off-diagonal elements
of the Hamiltonian eq.(\ref{HAMN1}) to higher n states yield corrections
of order $(\mu_B B / R_{\infty})^2$ and can be neglected for
magnetic field strengths of practical interest.\\
There has been recent activity in evaluating additional higher order individual contributions to the M energy levels, including in particular the ground state HFS\cite{Eides_2006,Karshenboim_2005,Jentschura_2005,Jentschura_2006,Karshenboim_2006,Eides_2007,Terekidi_2007,Karshenboim_2008,Eides_2009,Eides_2009a,Mondejar_2010,Eides_2010,Nuomura_2013,Eides_2013,Eides_2014,Eides_2014a,Eides_2014b,Eides_2014c}. The prospects for theory are that 10~Hz precision can be reached for $\Delta \nu_{HFS}$.\cite{Eides_2014} At this moment we are awaiting a new and fully coherent compilation of all to date calculated terms which enables an accurate comparison of theory values with a  measurable quantity. \\
\section{Muonium Production}
Polarized muons are available from intense muon channels at several accelerator centers. For precision M  experiments $\mu^+$ beams at surface muon momentum ($\approx$29~MeV/c) and below are particularly important. Such beams originate from pion decays at rest at (or near) the surface of a muon production target in an intense proton beam. The $\mu^+$ are highly polarized (up to 100\%) due to parity violation in the decay $\pi^+ \rightarrow \mu^+ + \nu_{\mu}$ from which $\mu^+$ are collected and transported in a beam line \cite{Bowen_1985}. 
Production of M atoms can occur by $e^-$ capture, when $\mu^+$ are stopped in matter \cite{Hughes_1977}. For precision fundamental experiments several different methods exist:\\
%\begin{itemize}
%\leftmargin=0mm
%\item
$(i)$ ~{M is formed after $\mu^+$ are stopped and thermalized in gases with suited electron binding energies such as N$_2$, Ar and Kr. The $\mu^+$ are slowed down by collisions with gas atoms. In the final stages of the stopping process frequent $e^-$ capture and re-stripping occur. In the end  M can be formed by $e^-$ capture. In inert gases, i.e. in such with ionization energies above 13.55~eV no further chemical reactions take place \cite{Hughes_1977}. Production yields for  cases of practical importance are 65(5)~\%for Ar, 80(10)~\% for Kr and 100~\% for Xe \cite{Schwarz_1993,Stambaugh_1974}.  
Due to collisional quenching the atoms are in the n=1 state. The muon polarization in the sample of M atoms depends on the magnitude of an external magnetic field B \cite{Thompson_1973}. 
In high B fields ($x\gg 1$), for  $\mu^+$ polarization P and
for unpolarized electrons in the target the four hyperfine states are populated with fractions $f_i$, $i=1,...,4$,  (see Fig. \ref{81403Fig2})
\begin{eqnarray}
\label{POL}
f_1= &\frac{1}{4} \left( 1+ P \right) ,
	    &f_2= \frac{1}{4} \left( 1+ P(s^2-c^2) \right), \nonumber \\
f_3= &\frac{1}{4} \left( 1- P \right) ,
	    &f_4= \frac{1}{4} \left(( 1+ P(c^2-s^2) \right),
\end{eqnarray}
where 
$s=\sin(\frac{1}{2} {\rm arc~cot} x)~
{\rm and}~ c=\cos(\frac{1}{2} {\rm arc~cot} x)$
for the field parameter x from eq. \ref{XPAR}.
In collisions with paramagnetic gases such as O$_2$ the muon polarization is destroyed. Contaminations of the noble gas targets with paramagnetic atoms need to be kept below the some 10~ppm level \cite{Schwarz_1993}. This M production method is the default for measuring the ground state hyperfine interval $\Delta \nu_{HFS}$ in M \cite{Liu_1999}.}\\ 
%
%\item
$(ii)$ ~{M in vacuum has been produced by shooting $\mu^+$ at keV energies through thin foils. In this case electron transfer can result in M atoms in the ground state \cite{Bolton_1981}, M atoms in an excited state \cite{Oram_1981,Kuang_1989}, including the metastable 2s state, and M$^-$ ions \cite{Kuang_1989}. The process is most probable, if the velocity of the incoming  $\mu^+$ matches the electron orbital velocity in the target atoms.   } \\
%
%\item
$(iii)$ ~{M in vacuum has been formed by stopping a $\mu^+$ beam in hot metal foils, e.g. tungsten, from where M emerges at thermal energies. The $\mu^+$ thermalize in the foil, diffuse to the surface, and leave the solid surface after capturing an $e^-$ from the solid \cite{Mills_1986}.  This method  has found applications in producing slow muon beams by photoionization of M \cite{Nagamine_1995,Matsuda_2003,Bakule_2008,Bakule_2009}. }\\
%
%\item
$(iv)$ ~{\label{powder_target} M in vacuum is formed, if $\mu^+$ are stopped near the surface of a SiO$_2$ fluffy powder target \cite{Beer_1986}. The M atoms diffuse to the surface and leave it with thermal velocities. For the atoms in vacuum  a significant polarization of 39(9)~\% has been found \cite{Woodle_1988}, which agrees with the expected maximum fraction of 50~\% according to eq. {\ref{POL}. The fluffiness of the powder is essential and the production targets appear to loose efficiency on a timescale of approximately one week. This is simmilar to the time scale that was found for H$_2$O molecules leaving the SiO$_2$ targets after they are placed into a vacuum chamber \cite{Grossmann_1995}. For typical $\mu^+$ beam momentum bite $\Delta p/ p = 5$~\% typical average  yields of up to  2-5~\% have been achieved in precision experiments \cite{Schwarz_1992}. For this the targets had to be replaced twice a week  \cite{Willmann_1999}.
SiO$_2$ powder targets have been employed for a number of precision experiments on M, including searches for M$\overline{\rm M}$ conversion \cite{Marshall_1982,Willmann_1999} and  laser spectroscopic measurements of $\Delta \nu_{1s2s}$ \cite{Chu_1988, Jungmann_1990,Maas_1994,Meyer_2000}.}\\
%
%\item
$(v)$ ~{M atoms have been produced  in SiO$_2$ aerogel targets \cite{Schwarz_1992}. A fraction of the atoms has been observed to leave the SiO$_2$ aerogel into the surrounding vacuum, if the stopping of the $\mu^+$ beam  occurs close to the target surface. The yield of M in vacuum has been found to increase for an increased area of the  SiO$_2$ target surface \cite{Schwarz_1992,Springer_1993}. Recently research on this topic has been intensified. A significantly enhanced fraction of M that leaves the sample has been reported for low density SiO$_2$ aerogels. For these samples the surface had been enhanced by drilling holes with depth comparable to the width of the $\mu^+$ stopping distribution and their separation and diameter of order M diffusion distance within $\tau_{\mu}$ \cite{Antognini_2012,Bakule_2013,Beer_2014,Nagamine_2014}. A yield of 38(4)\% was found for mesoporous silica\cite{Antognini_2012}}, which is one order of magnitude above the yields achieved for SiO$_2$ powder targets.\\
%
%\end{itemize}
%
The recent progress in the production of  M in vacuum from low density aerogels is very encouraging. These findings can be expected to boost new experiments for which  M atoms in vacuum are essential. Among those are improved laser spectroscopy of M \cite{Boshier_1996}, a search for M$\overline{\rm M}$ conversion with increased sensitivity \cite{Willmann_1997} and investigations of the gravitational interaction of M \cite{Kirch_2014,Kaplan_2013}.The proposed experiment at J-PARC for measureing the muon g-2 value also draws on M atom production from aerogels and their subsequent photoionization \cite{Mibe_2011,Saito_2012}.

\section{Ground State Hyperfine Structure}
%\subsection{LAMPF experiment}
%
\begin{figure}
 \vspace*{8mm}
 \includegraphics[width=8cm]{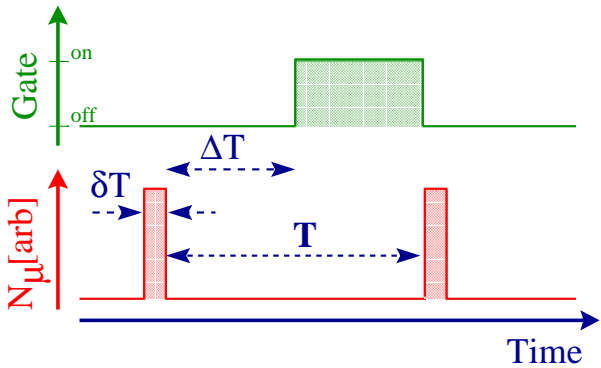}
 \vspace*{2mm}
 \caption{For the latest M hyperfine structure experiment at LAMPF the quasi continuous beam from the accelerator was modulated to produce $\delta$T = 4~$\mu$s long pulses with T=10$~\mu$s separation. Signals were recorded after a waiting time $\Delta$T. The produced M atoms interacted constantly and coherently with the microwave field in a cavity.}
 \label{81403Fig3}
\end{figure}
The HFS in the n=1 state in  M  has been measured in a series of experiments with increasing accuracy.
The latest and most accurate results were achieved by the Yale-Heidelberg-Syracuse collaboration at LAMPF. The experiment used the technique of {\it old muonium}
\cite{Liu_1999}. For this the atoms were produced with a pulsed $\mu^+$ beam and signals were recorded from M atoms that had lived and coherently interacted with an rf-field driving the HFS transitions significantly longer than the free muon lifetime $\tau_{\mu^+}$. This results in a resonance lineshape with a central feature that is narrower than the natural linewidth $\delta \nu_{HFS,nat} = 1 / (2 \pi \cdot \tau_{\mu^+}) = 144$kHz \cite{Boshier_1995}. This feature provides for a better determination of the line center. \\
\begin{figure}
 \vspace*{2.8cm}
 \hspace*{-7mm}
 \includegraphics[width=8.5cm]{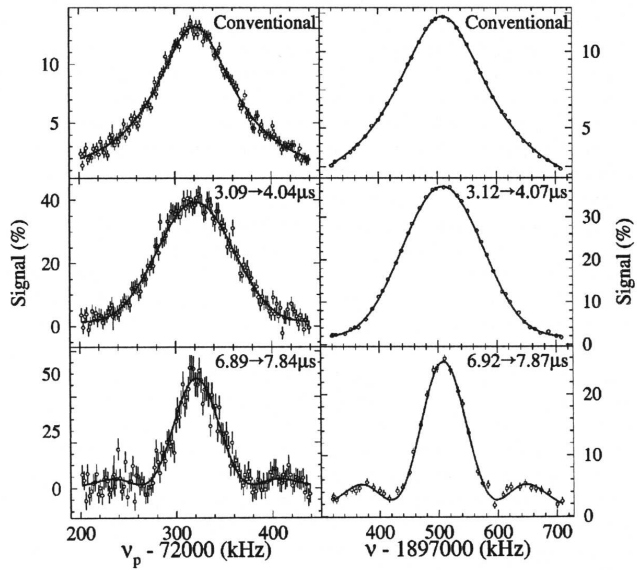}
 \vspace*{-2.2cm}
 \caption{Signals from the LAMPF M hyperfine structure experiment \cite{Liu_1999}. On the left are resonances obtained by sweeping the    magnetic field with microwave frequency kept constant, on the right are signals from sweeping the microwave frequency at constant magnetic field. The top row shows conventional microwave resonances, the second and third row correspond to {\it old muonium} resonances with the delay time indicated.}
\label{81403Fig4}
\end{figure}
In order to obtain a time structure suited for an {\it old muonium} experiment the cw surface $\mu^+$ beam at LAMPF  was chopped using an electrostatic deflector. The time structure is shown in Fig. \ref{81403Fig3}.
The beam had an average particle rate of some $10^7 \mu^+/s$ with close to 100~\% polarization and $\Delta p/p = 10\%$ momentum bite. It was stopped in a Kr gas target to produce M. The O$_2$ contamination in the target was minimized and kept below 5~ppm by circulating the gas constantly through a purifying system. The gas temperature was kept stable to 
$0.1~^{\circ}$C and its pressure was measured to 0.5~mbar. Data taking took place at $p_1$=800~mbar and $p_2$=1500~mbar. The lowest possible pressure was determined by the
stopping distribution inside the microwave cavity and had been chosen to keep the fraction of muons stopping in the walls of the cavity low.\\ 
The experiment was conducted in the homogeneous field of a large bore superconducting Magnetic Resonance Imaging (MRI) magnet at 1.7~T field. The magnet was equipped with a warm coil that provided for magnetic field modulation.  The microwave cavity had a quality factor around 25 000 and it was operated in  TM$_{110}$ mode for transitions at frequency $\Delta \nu_{12}$ and TM$_{210}$ mode for transitions at $\Delta \nu_{34}$ (Fig. \ref{81403Fig2}). Its resonance frequencies could be tuned with a movable quartz tuning bar.\\
For the measurements magnetic field scans as well as  microwave frequency sweeps were conducted in order to enable cross checks for various systematics. In both cases conventional and narrowed spectra were recorded (Fig. \ref{81403Fig4}). After correction for a small quadratic pressure shift at the pressures $p_1$ and $p_2$ an extrapolation to zero pressure was performed.
The magnetic field $B$ had been calibrated and monitored with the very same magnetic field measurement concept and devices \cite{Prigl_1996,Fei_1997} which also have been employed in the measurement of the muon magnetic anomaly $a_{\mu}$ at BNL \cite{Bennett_2006}.\\
As a consequence of eq. \ref{ZEEMAN} we have
\begin{eqnarray}
\hspace*{-5mm} \nu_{12} + \nu_{34} &=& \Delta \nu_{HFS}~~~ {\rm and}\\
\hspace*{-5mm} \nu_{12} - \nu_{34} &=& \frac{2~\mu_{\mu}~g_{\mu}`~B}{h}
	+ \Delta \nu_{HFS}
    \left[  (1+x^2)^{1/2}-x
    \right]~.
\end{eqnarray} 
With this the LAMPF experiment yielded \cite{Liu_1999}
\begin{eqnarray}
\Delta \nu_{HFS} = 4\,463\,302\,776(51)~{\rm Hz}~(11~{\rm ppb})~~~~, \\
\mu_{\mu} = 3.183\,345\,24(37)~~~~(120~{\rm ppb})~~~~, \label{mumu}\\
\frac{m_{\mu}}{m_e} = 206.768\,277(24)~~~~(120~{\rm ppb})~~~~.
\end{eqnarray} 
An accurate value for the fine structure constant was extracted as
$ \alpha^{-1}$=$137.035\,996\,3(80)~(58~{\rm ppb})$.
The good agreement of this result with the most precise value from the electron magnetic anomaly \cite{Aoyama_2012,Hanneke_2011} 
$ \alpha^{-1}$=$137.035\,999\,173(35)~(0.26~{\rm ppb})$
is a test
of internal consistency of QED, because one involves theory of bound states and the other one of free particles.

The LAMPF experiment has been analyzed for a potential sidereal variations of the measured frequencies. No signal was found and thereby stringent limits for second generation particles could be set \cite{Hughes_2001}.
Based on a Standard Model extension model, further stringent tests of Lorentz and CPT invariance in muon physics have been suggested \cite{Gomes_2014}. This theoretical framework has no predictive power and it involves a multitude of potential parameters, which all could indicate potential Lorentz and CPT violation. However, it provides a robust theoretical framework for a quantitative evaluation and comparison of all experiments in the area.\\

%\subsection{J-PARC experiment}

From  earlier measurements in gases \cite{Mariam_1982,Casperson_1977} the pressure dependence of $\Delta \nu_{HFS}$  at pressures $p$ up to about 100~bar has been reported as 
\begin{equation}
\Delta \nu_{HFS}(p) = (1+ap+bp^2) \cdot \Delta \nu_{HFS}~~,
\end{equation}
where $\Delta \nu_{HFS}$ is the vacuum value. For Kr the coefficients
are $a_{Kr}=7,996(8)\cdot10^{-6}/$~bar and $b_{Kr}=5.5(1.1)\cdot10^{-9}/$~bar$^2$. 
The term quadratic in $p$ has been verified by fitting data obtained in measurements at pressures up to about 100 bar. For $p$=1~bar the quadratic shift amounts to 25(5)~Hz. This calls for attention in the next round of precision experiments which aim to improve the LAMPF result by one order of magnitude \cite{Torii_2015,Shimomura_2015}. In particular, experiments which determine $\Delta \nu_{HFS}$ have been conducted all at gas pressures where many-body collisions dominate. Note, in the regime, where two-body collision dominate, deviations from mostly linear behaviour have been observed for, e.g. atomic H, and hyperfine decoupling could be established at $p<1$~mbar \cite{Weber_1979}.\\
At J-PARC a new experiment to measure $\Delta \nu_{HFS}$ and $\mu_{\mu}$ is in progress \cite{Torii_2015,Shimomura_2015}. It uses a concept  similar  the latest LAMPF experiment, however, with significantly improved technology. In particular, a new and intense beam line \cite{Sasaki_2013} will be employed. The microwave cavity will have an  extended length to reduce $\mu^+$ stopping in the cavity walls and to enable measurements at lower gas pressures. New $B$ field calibration by proton NMR will be used and $^3$He based field calibration is considered. The project is well underway and aims for gaining one order of magnitude over the latest measurement. A new directly measured value for $\mu_{\mu}$ is urgently needed. The new experiments to measure $a_{\mu}$ require such a value which is independent of elaborate QED theory and which has not been extracted from $\Delta \nu_{HFS}$. Note, here QED theory is exploited to extract $\mu_{\mu}$, the theory which is stringently tested by $a_{\mu}$. %The two values agree within the accuracy eq. \ref{mumu}.
\\

%\subsection{New opportunities}
%
\begin{figure}
\vspace*{6mm}
 \includegraphics[width=8cm]{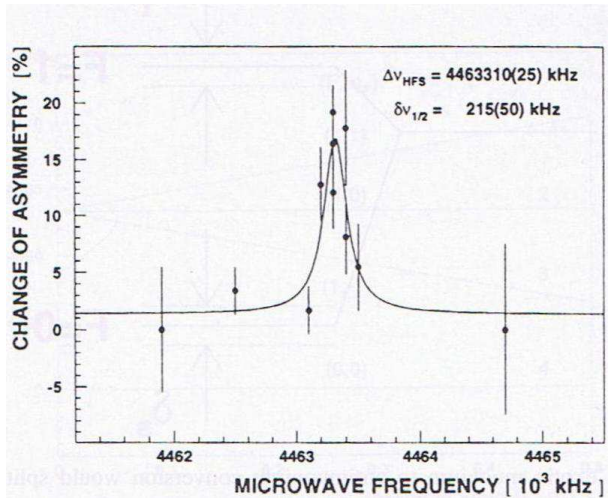}
\vspace*{0mm}
 \caption{Signal of M ground state hyperfine structure in vacuum\cite{Jungmann_1995}. Thermal M atoms from a SiO$_2$ powder target emerged into a microwave cavity where they interacted with a microwave field. }
 \label{81403Fig5}
\end{figure}
The observation of polarized M in vacuum \cite{Woodle_1988} has also enabled spectroscopy
of n=1 state HFS transitions in vacuum. 
An advantage arises from the absence of buffer gas related corrections in such an experiment. 
A first experiment has been conducted using a cw $\mu^+$ beam at the Paul Scherrer Institut (PSI), Villigen, Switzerland, \cite{Jungmann_1995} 
to measure $\Delta \nu_{HFS}$.
An obstacle to overcome is the finite average velocity of the atoms. Whereas in a gas the atoms effectively 
do not move, atoms emerging from a solid state surface have thermal velocity distribution. In the experiment the average velocity
was ${v}_{M,therm}=7.4(1)$~mm/$\mu$s \cite{Woodle_1988}, 
which corresponds to T = 296(10)~K temperature. The atoms have in vacuum an average range of 16.3(2)~mm. 
They had 39(9)~\% polarization and were entering a microwave cavity located in 
less than 0.6~$\mu$T background magnetic field. Inside the cavity transitions could be induced between the 1$^2$S$_{1/2}$,F=1,m$_F$=1 
and 1$^2$S$_{1/2}$,F=1,m$_F$=0 levels. They were signaled by a change of the $\mu^+$ 
decay asymmetry (see Fig. \ref{81403Fig5}). The resonance frequency was determined as 
$\Delta \nu_{HFS}$ = 1\,463\,310(25)~kHz. The signal width was $\delta \nu_{HFS}$ = 215(50)~kHz,
which is consistent with the natural linewidth $\delta \nu_{nat}$ = 144~kHz, microwave power broadening  
$\delta \nu_{HFS,power}$= 85(11)~kHz and Doppler broadening $\delta \nu_D$ = 45(10)~kHz.\\ 
At an intense pulsed $\mu^+$ source and with  modern decay positron tracing, the Doppler 
effect could be significantly reduced, in particular, if the experiment were conducted with the {\it old muonium} technique. 
Since studies for pressure shifts are obsolete, an experiment on $\Delta \nu_{HFS}$ in vacuum requires less $\mu^+$ from a beam 
than experiments which use $\mu^+$ stopping in a gas.\\

\section{1s-2s Transition}
\begin{figure}
 \vspace*{3mm}
 \includegraphics[width=9.0cm]{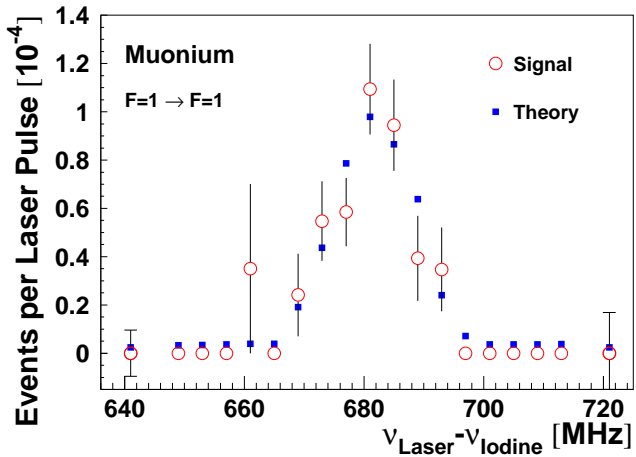}
 \vspace*{0mm}
 \caption{The 1$^2$S$_{1/2}$,F=1 - 2$^2S${1/2},F=1 transition in M\cite{Meyer_2000}. The transition was observed through subsequent photo-ionization of the excited state and particle detection of the released $\mu^+$.}
 \label{81403Fig6}
\end{figure}
The gross structure splitting $\Delta \nu_{1s2s}$ in M has was measured in a pioneering experiment \cite{Chu_1988,Danzmann_1989}
at the High Energy Accelerator Research Organization (KEK), Tsukuba, Japan, and in a series of measurements with increasing accuracy \cite{Jungmann_1991,Maas_1994,Meyer_2000} at the Rutherford Appleton Laboratory (RAL), Chilton, U.K., using Doppler-free two-photon laser spectroscopy. M atoms were produced from SiO$_2$ powder targets (see sec.\ref{powder_target}.).
Laser induced two-photon transitions 1$^2$S$_{1/2}$,F=1-2$^2$S$_{1/2}$,F=1 could be detected via photo-ionization of the excited state by a third photon from the same laser field and the subsequent particle detection of the released $\mu^+$. 
 The early experiments used
excimer laser pumped dye lasers with pulse lengths of order 15~ns \cite{Chu_1988,Jungmann_1991}. Their light was frequency doubled in an electro-optical crystal to obtain light at wavelength 244~nm (see Fig. \ref{81403Fig1}). Doppler-free two-photon signals were achieved by retro-reflecting the laser beam onto itself. At the high pulsed laser intensities frequency chirping occurs at several 10~MHz due to rapid optical phase changes in the pulsed optical amplifiers \cite{Reinhard_1996}. This effect limits the accuracy to which the transition frequencies can be measured with fast pulsed laser systems. \\
The most recent experiment employed a cw dye laser seeded pulsed alexandrite laser amplifier at wavelength 732~nm the light of which was frequency tripled. The frequency chirp of the alexandrite laser amplifier during  100~ns long pulses could be compensated to below 10~MHz \cite{Bakule_2000}.  The chirp was measured pulse by pulse \cite{Reinhard_1996,Meyer_2000} and it was correlated with individual observed photoionization events. The signal analysis was based on a line shape model \cite{Yakhontov_1996,Yakhontov_1999} which takes chirped excitation into account (Fig. \ref{81403Fig6}). Calibrated
lines in molecular I$_2$ \cite{Bagayev_2000,Cornish_2000}
served as frequency reference. 
The experiment yielded \cite{Meyer_2000}
\begin{equation}
 \Delta \nu_{1s2s}  = 2\,455\,528\,941.0(9.8)~{\rm MHz}~~,
\end{equation}
with a Lambshift contribution of $\Delta \nu_{LS} = 7\,049.4(4.9)~{\rm MHz}$.
The $\mu^+$ to $e^-$ mass ratio was extracted as
\begin{equation}
\frac{m_{\mu^+}}{m_{e^-}} = 206.768\,38(17)~~. 
\end{equation}
This agrees with the value from $\Delta \nu_{HFS}$ in M \cite{Liu_1999}
\begin{equation}
\frac{m_{\mu^+}}{m_{e^-}} = 206.768\,277(24)~~.
\end{equation}
The measured isotope shift between M and D has been improved recently with a recalibration of
the relevant  molecular I$_2$ frequency reference. It is now \cite{Fan_2014} 
%\begin{equation}
$ \Delta \nu_{1s2s}(M-D)  = 11\,203\,464.9(10)~{\rm MHz}$,
%\end{equation}
in agreement with theory.\\

Frequency chirping in the pulsed laser amplifiers \cite{Reinhard_1996} were the main systematics in the experiments on $\Delta \nu_{1s2s}$ to date.
Therefore, significant progress can be expected only from future experiments with cw laser systems, where frequency chirps are absent.
The intensity of cw laser light can be enhanced significantly in an optical cavity. Effective photoionization for the detection of the excited state requires an intense second light field. For this step resonance enhancement can be exploited. 
Along these lines an experiment has been proposed  \cite{Mills_2014} which has a potential to achieve almost $\Delta \nu_{nat}$ linewidth. The atoms are produced with a pulsed $\mu^+$ beam from porous silica targets. They interact in a suited enhancement cavity for about $\tau_M$ with a standing wave light field at 244~nm wavelength. The n=2 state population is detected via excitation to the 15P state with light at 368~nm wavelength and subsequent photoionization of the atom and detection of the $\mu^+$. An event rate of $10^4$ per day is predicted. Such an experiment is expected to provide for a measurement of $\Delta \nu_{1s2s}$ to  about 1~kHz ($\approx$10ppt). With such precision the knowledge of $m_{\mu}$ can be improved over the presently best value \cite{Mohr_2012} by one order of magnitude.  \\

%\subsection{New possibilities}

\section{Future Possibilities}
%\subsection{Muonium-Antimuonium conversion}
The increased M production yields reported for SiO$_2$ aerogel targets together with the upcoming intense $\mu^+$ sources suggest also to consider a  new search for M$\overline{\rm M}$ conversion. For neutral kaons K$^0$$\overline{\rm K^0}$ oscillations are a well established reality in the hadron sector. In the lepton sector the process involving the leptons e$^-$ and $\mu^+$, which correspond to the quarks in  K$^0$ and $\overline{\rm K^0}$, has not been observed, yet. The last experiment \cite{Willmann_1999} was statistics limited by the available $\mu^+$ flux. For a new experiment 2 to 3 orders of magnitude higher sensitivity is within reach through exploiting the time evolution of the  M$\overline{\rm M}$ conversion process (Fig. \ref{81403Fig7}). For this, at a pulsed $\mu^+$ source the ratio of the potential $\overline{\rm M}$ signal to $\mu^+$ and M decay related background could be significantly
reduced, in particular if a N-fold coincidence signal signature is employed.
In this case the muon related background is  expected to drop exponentially with a time constant 
$\tau_{\mu}/{\rm N}= \lambda_{\mu}^{-1}/{\rm N}$~ \cite{Willmann_1997}.\\
%
%\subsection{Search for antigravity}
A clear status assignment {\it matter} or {\it antimatter} is not obvious for the M atom. Firstly, the $\mu^+$ is an antiparticle. With ${m_\mu} \approx 207 m_e$ potential antigravity can be expected to dominate for  M. Secondly, as far as lepton numbers are concerned, we have one particle and one antiparticle which puts the status of M right in between such as it is the case also for positronium. At present we do not have any proven theory that can reliably rule out antigravity for M. The question needs to be investigated experimentally. For this, recently a new idea has been proposed at PSI to search for potential antigravity using a Mach Zehender interferometer \cite{Kirch_2014,Kaplan_2013}. With intense muon beams sufficient statistics can be expected to find a potential difference in the free fall component of free moving M atoms. Such an experiment is particularly interesting, because its interpretation is expected to be rather straightforward, unlike comparable experiments with protons and antiprotons, which are in progress at the CERN AD facility. For protons and anti-protons the inner structure of the particles contains next to three quarks and three antiquarks primarily gluons, the energy of which is responsible for most of the measured mass. They are the same in protons and antiprotons which lets us expect at most a small difference in the behaviour and all properties of both particles at all\cite{Willmann_2015}. One would therefore rather expect a small difference in the hadron and antihadron gravitational interaction. \\
\begin{figure}
\vspace*{5.3cm}
 \hspace*{-0.6cm}
 \includegraphics[,width=8.3cm]{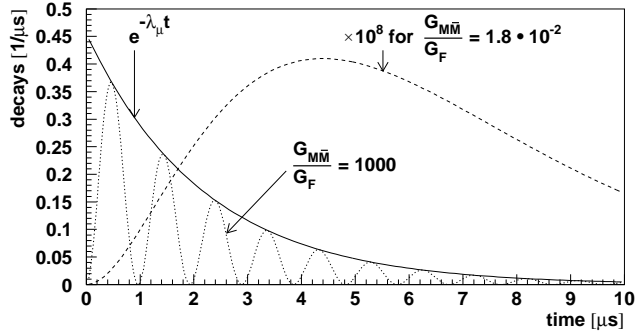}
 \vspace*{-4.6cm} 
 \caption{The probability to observe an $\overline{\rm M}$ decay increases with time to a maximum after some $2 \tau_{\mu}$. The ratio of $\overline{\rm M}$ over $\mu^+$ decays increases further towards later times. This provides for a possible  enhanced signal to background ratio in an experiment where the time between M formation and  M or $\overline{\rm M}$ decay is recorded \cite{Willmann_1997}.
 \label{81403Fig7}}
\end{figure}
%
%\subsection{Exotic searches}
%
Next to the decay of M via Standard Model allowed  $\mu^+$ decays and M$\overline{\rm M}$ conversion, M  could decay also via  a rare muon decay. Among the possible decays of the atom its possible disintegration into  4 neutrinos has been theoretically studied \cite{Gninenko_2013}. Yet, present and near future source strengths at existing muon facilities as well as achievable detector efficiencies prohibit experiments. More realistic is the exploitation of precision spectroscopy experiments for dark matter searches. Here already constraints could be imposed recently from the results of HFS measurements \cite{Karshenboim_2014}.

\section{Conclusions}
{\it Muonium has not yet decayed} is the title of an article by Hughes and zu Putlitz in 1984 \cite{Hughes_1984}. The statement is still true today, because at now available intense muon sources such as at J-PARC precision experiments can enter a next level of accuracy, where the Standard Model theory can be stringently tested, important fundamental constants \cite{Mohr_2012} can be accurately measured, and powerful searches for new and exotic interactions can be conducted. Every forefront precision experiment puts the hypothesis of lepton universality to test.\\

%\begin{acknowledgment}
\acknowledgment
The author would like to thank the Dutch funding agency FOM for partial support of this work.
%\end{acknowledgment}

% \bibliography{jpsj_KJ}

%
\profile{Klaus P. Jungmann} {was born in Heidelberg, Germany, in 1955. He received a diploma in Physics in 1981 and a doctoral degree in 1985 from the University of Heidelberg, Germny. Between 1985-1987 was postdoc at the IBM Almaden Research Center in San Jose, California, U.S.A. From 1987-2001 he was at the Physikalisches Institut at the University of Heidelberg as postdoctoral researcher, research assistant, University Docent and Extraordinary Professor of Physics. From there he conducted precision experiments in muon physics at the Paul Scherrer Institute, Switzerland, the Rutherford-Appleton Laboratory, United Kingdom, the Los Alamos National Laboratory, U.S.A., and the Brookhaven National Laboratory, U.S.A. This included a search for muonium-antimuonium conversion, laser spectroscopy of muonium , and  measurements of the muonium hyperfine structure and the muon magnetic anomaly. Since 2001 he is professor at the University of Groningen, The Netherlands, where he leads precision experiments on trapped radioactive and stable isotopes.
%, currently addressing parity violation and permanent electric dipole moments.
}  
%
%\vspace*{2cm}
%\begin{figure}[hbt]
%\vspace*{5.3cm}
% \hspace*{-0.6cm}
% \includegraphics[,width=8.3cm]{81403Fig8.eps}
%\vspace*{-4.6cm} 
 %\caption{Klaus Jungmann.
% \label{81403Fig8}}
%\end{figure}
%
   
\end{document}